\documentclass[prb,twocolumn,superscriptaddress,showpacs,aps,footinbib,10pt]{revtex4-1}

\usepackage{color}
\usepackage{amsmath}
\usepackage{graphicx}
\usepackage{amssymb}
\usepackage{amsthm, amscd}
\usepackage{amsfonts}
\usepackage{times}
\usepackage[abs]{overpic}
\usepackage[english]{babel}
\usepackage[colorinlistoftodos]{todonotes}

\newcommand{\be}{\begin{eqnarray}}
\newcommand{\ee}{\end{eqnarray}}
\newcommand{\beq}{\begin{eqnarray}}
\newcommand{\eeq}{\end{eqnarray}}
\newcommand{\ket}[1]{\left|{#1}\right\rangle}
\newcommand{\bra}[1]{\left\langle{#1}\right|}
\newcommand{\up}{\uparrow}
\newcommand{\dn}{\downarrow}
\newcommand{\br}{\mathbf{r}}
\newcommand{\bR}{\mathbf{R}}

\newcommand{\bk}{\mathbf{k}}

\begin{document}

\title{Fractional Chern insulators in bands with zero Berry curvature}

\author{Steven H. Simon}
\author{Fenner Harper}
\affiliation{Rudolf Peierls Centre for Theoretical Physics, University of Oxford, Oxford, OX1 3NP, United Kingdom}

\author{N. Read}
\affiliation{Department of Physics, Yale University, P.O. Box 208120, New Haven, Connecticut 06520-8120, USA}

\date{\today}

\begin{abstract}
Even if a noninteracting system has zero Berry curvature everywhere in the Brillouin zone, it is possible to introduce interactions that stabilize a fractional Chern insulator. These interactions necessarily break time-reversal symmetry (either spontaneously or explicitly) and have the effect of altering the underlying band structure. We outline a number of ways in which this may be achieved, and show how similar interactions may also be used to create a (time-reversal symmetric) fractional topological insulator.   While our approach is rigorous in the limit of long-range interactions, we show numerically that even for short-range interactions a fractional Chern insulator can be stabilized in a band with zero Berry curvature. 
\end{abstract}

\maketitle

\section{Introduction}
There has been a great deal of recent interest in fractional Chern insulators (FCIs), which are the lattice analogs of fractional quantum Hall (FQH) states.\cite{FCIreviews} In addition to providing new experimental settings in which to explore topologically ordered phases of matter, FCIs demonstrate lattice effects that substantially change the underlying physics from the quantum Hall picture.

The usual description of FCIs proceeds in the following way: Imagine starting with a so-called Chern band---a band which, when completely filled with noninteracting electrons would have nonzero quantized Hall conductivity. One next imagines partially filling the band and introducing an interaction which creates a gapped, many-body ground state. If the interaction is chosen correctly, the resulting state will have a fractionally quantized Hall conductivity.  

While this has proved to be a reasonable and successful approach to the problem, we believe it overemphasizes the structure of the noninteracting band, which may be irrelevant if the particles interact strongly. To make this point clear, we will construct an example that starts with a noninteracting system having zero Berry curvature everywhere in the Brillouin zone. Then, by adding an appropriately chosen interaction that breaks time reversal symmetry (either explicitly or spontaneously), we will find that the ground state is an FCI. We will then show how similar interactions may be used to generate a time reversal-symmetric, fractional quantum spin Hall insulator,\cite{KaneMele} which is an example of a fractional topological insulator (FTI).   
We support our findings with numerical evidence from exact diagonalization studies.  While our analytic arguments rely on the use of very long-range interactions, we find in our numerical work that even fairly short range interactions can form fractional Chern insulators in bands with zero Berry curvature.

\section{FCI in a band with Nonzero Chern Number \label{nonzerocham}}
We will first outline the conventional method for obtaining an FCI state from an underlying topological band structure. In Sec.~\ref{interacting_chern} we will then attempt to reproduce this state, starting from a band with zero Berry curvature.

Although our construction is very general, let us restrict our attention to the case of spinless particles on a honeycomb. This is just a triangular lattice with two sites per unit cell, which we will call $A$ and $B$ sites. We consider a kinetic energy,
\begin{eqnarray}
  \tilde   K &=& \sum_{\br,\br'} \tilde{t}_{\br\br'} \, c^\dagger_{\br} c^{\phantom{\dagger}}_{\br'},\label{chern_kinetic}
\end{eqnarray}
where $\br$ and $\br'$ are summed over the sites of the lattice and $\tilde t_{\br\br'}$ describes the hopping amplitudes. In order to produce a Chern band, the hoppings (which are short-range) couple the sublattices, and phases are introduced which break time reversal symmetry. Typically one tunes hoppings up to third nearest neighbors in order to get a very flat (although not perfectly flat) Chern band.  If desired, one can use further neighbor hoppings to make the bands extremely flat. For example, $\tilde{t}_{\br\br'}$ may be chosen to describe the flattened Haldane honeycomb model.\cite{HaldaneHoneycomb,Sheng,NeupertPRL2011}

If one introduces a suitable short-range interaction (such as a nearest-neighbor repulsion), which we write as
\be
 \tilde V &=& \sum_{\br\br'}   v_{\br\br'} \hat{n}_{\br}  \hat{n}_{\br'},
\ee
with $\hat{n}_{\br} = c^\dagger_{\br} c^{\phantom{\dagger}}_{\br}$ being the particle number on site $\br$, then the total Hamiltonian
\begin{equation}
\label{eq:Ham1}
  \tilde H = \tilde K + \tilde V
  \end{equation}
will produce an FCI ground state at the appropriate particle-filling fraction. For example, FCI states analogous to Laughlin states may be observed at $\nu=1/3$ for fermions and $\nu=1/2$ for bosons.  This conclusion has been established numerically by a number of groups.\cite{Regnault,Sheng}

\section{FCI in a Band with Zero Berry Curvature\label{interacting_chern}}
Now consider a different band structure, described by a kinetic energy $K$, which has zero Berry curvature throughout the Brillouin zone. We claim that we can switch on an interaction, $V$, such that the ground state of this new model is also an FCI and is adiabatically connected to the FCI described in Sec.~\ref{nonzerocham}. Our plan is to choose the interaction $V$ such that $H=K+V$ mimics the Hamiltonian~\eqref{eq:Ham1}, which is known to give an FCI.

To be specific, let us write the noninteracting kinetic energy as
\be
  K &=& \sum_{\br, \br'} t_{\br\br'} \, c^\dagger_{\br} c^{\phantom{\dagger}}_{\br'},
\ee
where $t_{\br\br'}$ describes new hopping parameters. For simplicity we will consider a hopping model which is diagonal in the sublattices, i.e.,
\be
 t_{\br,\br'} &=&\left\{
 \renewcommand\arraystretch{1.5}
 \begin{array}{cc}
 t^A_{\br\br'} & \mbox{if $\br,\br'\in A$}, \\
 t^B_{\br\br'} & \mbox{if $\br,\br'\in B$}, \\
 0& \mbox{otherwise.}
 \end{array}
 \right.\label{eq:t_def}
 \ee
We also break the symmetry between sublattices by giving each one a different on-site energy $t_{\br\br}$. In this way, we can arrange such that the two bands do not overlap in energy, the eigenstates are of the simple plane wave form $e^{i {\bf k \cdot R}}$, and each band involves only one sublattice. Since the bands are completely decoupled, for any translationally invariant hopping $t_{\br\br'}$, the eigenfunctions have trivial structure and all bands have zero Berry curvature throughout the Brillouin zone (see Appendix~\ref{berry_def} for the definition of this quantity).

Now, certain interactions, which are short-range with a decay length $\xi$ on the order of several lattice spacings, can be treated accurately within mean-field theory as long as $\xi$ is much larger than the interparticle spacing. Consider an operator 
\begin{eqnarray}
 \hat N_\br =\frac{1}{\bar n} \sum_{\br'}  f(\br' - \br)  \hat{n}_{\br'}\label{eq:loc_dens}
\end{eqnarray}
where $\bar n$ is the average density of electrons in the system. In this expression, $f$ is some function (like a Gaussian) that is smooth and slowly decaying with length scale $\xi$, and has normalization   
\be
\sum_{\br'}  f(\br' - \br) = 1.\label{eq:f_normalization}
\ee
By short-range with decay length $\xi$, we mean that the function $f(\br'-\br)$ satisfies 
\be
\left|f(\br'-\br)\right|<Ce^{-|\br'-\br|/\xi}
\ee
for all sufficiently large $|\br'-\br|$ and with finite constants $C$ and $\xi$.

In this way, the operator $\hat N_\br$ measures the density of electrons in a region of length scale $\xi$ and compares it to the average density, $\bar{n}$. If the state being operated on has approximately uniform density, and if we take $\xi$ to be large enough, it must be the case that $\hat N_\br$ is extremely close to unity (since we are comparing the average density in a very large region to the real average density). We can then accurately approximate $\hat N_\br$ by unity and do not need to treat it as an operator.  As we take $\xi$ larger and larger, this approximation becomes more and more accurate (assuming there is no phase separation). The fluctuations of $\hat N_\br$ around this average may be treated perturbatively (see Appendix~\ref{pert_stab}). 

We may now consider the interaction term
\be
  U = \frac{1}{2}\sum_{\br\br'}   T_{\br\br'} :\left[\hat N_\br c^\dagger_{\br} c^{\phantom{\dagger}}_{\br'}+c^\dagger_{\br} c^{\phantom{\dagger}}_{\br'} \hat N_{\br'}\right]:,\label{eq:hopping_interaction}
\ee
where $T_{\br\br'}$ is an interaction strength (to be described below) and the entire expression is normal ordered. This operator is Hermitian and includes four-fermion interaction operators, although it is not generally a density-density interaction. The interaction is short-range in that it involves particles that are separated by a maximum distance on the order of $\xi$.  In the mean-field limit, $\hat N_\br$ becomes unity and we obtain
\be
  U \rightarrow   \sum_{\br\br'} T_{\br\br'} \,c^\dagger_{\br} c^{\phantom{\dagger}}_{\br'},
\ee
which is simply a hopping term. Thus, let us choose
\be
  T_{\br\br'} = \tilde{t}_{\br\br'}  - t_{\br\br'},
\ee
where $t_{\br\br'}$ and $\tilde{t}_{\br\br'}$ were defined in Eqs.~\eqref{eq:t_def} and \eqref{chern_kinetic}, respectively, so that  
\be
  K + U  \rightarrow \tilde K
\ee
at the mean-field level. In this way, our scheme uses an interaction to modify the ``effective" band structure of the underlying (topologically trivial) model. Note that since $\tilde t_{\br\br'}$ breaks time reversal, so too does $T_{\br\br'}$.\cite{ordering_footnote}

One might worry that this density-dependent interaction is unstable to phase separation, and indeed, it is fairly simple to write down an expression for $T_{\br\br'}$ for which this occurs. To make sure that this is always disfavored, we can add a diagonal term to the interaction strength, writing
\begin{eqnarray}
T_{\br\br'}&\to&T_{\br\br'}+M\delta_{\br\br'} \label{modified_T},
\end{eqnarray}
where the second term may be absorbed into the amplitude $T_{\br\br'}$. When added to Eq.~\eqref{eq:hopping_interaction}, this diagonal term generates a density-density interaction between a particle at $\br$ and its surrounding neighborhood, suppressing density fluctuations on the scale of $\xi$. By adjusting the strength of $M$, we can ensure that complete phase separation is never favored, but that the small density fluctuations required for an FCI state are allowed.  

As mentioned previously, our mean-field picture is exact in the large $\xi$ limit. Another concern is whether this mean-field state remains stable in the presence of fluctuations about this limit. In Appendix~\ref{pert_stab} we discuss the perturbative corrections due to the interaction term $U$ and justify the stability of the FCI state. We then provide numerical evidence in support of the FCI state in Sec.~\ref{numerics}.

The interaction $U$ changes the effective band structure of the model, so if we now use the total interaction 
\be
 V = \tilde V + U,
\ee
then the Hamiltonian $H = K + V$ completely mimics the above $\tilde H$ [Eq. \eqref{eq:Ham1}] and will have an FCI ground state. As required, the noninteracting band structure of $K$ has zero Berry curvature and the topological properties are induced purely by the interaction term, $V$.

The idea underpinning this approach is that hopping terms are essentially a subset of all possible interaction terms. By introducing an auxiliary operator, in this case a local density operator, that acts as a number on the ground state, an interaction can be written down that modifies the effective single-particle band structure. The trade-off for this is that the interactions are no longer density-density, and have a range that is finite but may cover several unit-cell lengths.

In the expression above we have broken time-reversal symmetry by hand with our choice of interaction $U$. If preferred, it is possible to generate an FCI state spontaneously with a time-reversal-invariant interaction, as discussed in the next section. This requires us to introduce a spin-1/2 degree of freedom but will also allow quantum spin Hall insulator (or FTI) states to be realized.

\section{Time-Reversal-Invariant Interactions}
\subsection{Spin-dependent band models}
One objection to the approach outlined above is that it breaks time-reversal symmetry explicitly through the interaction $U$. In order to use a time-reversal-symmetric interaction, we now promote the spinless fermions to spin-$1/2$ fermions. These obey the usual interrelations under time reversal, which can be written as
\be
\Theta c_{\br\sigma}\Theta^{-1}&=&i\tau^{y}_{\sigma\sigma'}c_{\br\sigma'},
\ee
where $\tau^y$ is the Pauli $y$-matrix, $\sigma$ and $\sigma'$ are spin labels, and $\Theta$ is the time reversal operator.\cite{BernevigBook} We have also included a spin index on the field operators $c_{\br\sigma}$.

The underlying topologically trivial bands must now have a spin dependence. For simplicity, we will treat spin as a good quantum number and write the kinetic-energy term as
\be
  K' &=& \sum_{\br\br'\sigma} t'_{\br\br'\sigma} \, c^\dagger_{\br\sigma} c^{\phantom{\dagger}}_{\br'\sigma},
\ee
although in general we could have included spin-flipping terms. As before, we will assume that the different sublattices have different on-site energies. Then, following the arguments of Sec.~\ref{interacting_chern}, each of the four bands will have exactly zero Berry curvature.

For the hopping parameters of our target Chern band model, we take $\tilde{t}_{\br\br'}$ from Eq.~\eqref{chern_kinetic} for one spin species but, crucially, invert the Chern number for the other spin species by taking the conjugate hopping. In other words, we take
\be
\tilde{t}'_{\br\br'\up\up} &=&\tilde{t}_{\br\br'}, \\
\tilde{t}'_{\br\br'\dn\dn} &=&\left[\tilde{t}_{\br\br'} \right]^*
\ee
and set any spin-flipping terms to zero. This defines a new topologically nontrivial kinetic energy, $\tilde{K}'$, which one can easily show is symmetric under time reversal symmetry. This kinetic energy therefore generates a low-energy band for each spin species, and these have equal and opposite (nonzero) Chern numbers when filled.

The framework outlined above leads to a Hamiltonian that conserves spin. For example, $\tilde{K}'$ could describe the kinetic energy of the Kane-Mele model \emph{without} the Rashba spin-orbit term.\cite{KaneMele} In principle, our approach could be generalized to time-reversal invariant models that do not conserve spin, such as the full Kane-Mele model. However, in that case, the interactions we describe below would involve complicated terms that mix both physical spin species; we will restrict our discussion to the simpler, spin-conserving case.

To generate a fractional state, we would partially fill the lowest band for each spin species and turn on a short-range interaction,
\be
\tilde{V}'= \sum_{\br\br'\sigma}   v'_{\br\br'\sigma} \hat{n}_{\br\sigma}  \hat{n}_{\br'\sigma}.
\ee
This is diagonal in the spin index, so at this stage there is no interaction between spins. 

The total Hamiltonian
\begin{eqnarray}
\tilde{H}'&=&\tilde{K}'+\tilde{V}'\label{eq:HamTR}
\end{eqnarray}
would generate a fractional topological state, but the overall Chern and spin Chern numbers would depend on the filling fraction of each of the two lowest spin bands. If both bands are filled equally, say at $\nu=1/3$ each, then the system will retain its time-reversal symmetry and will form a fractional topological insulator state. If one band remains empty but the other is (fractionally) filled, then the system could spontaneously break time-reversal symmetry and form an FCI state, depending on the interaction $V$. The relative stability of these two possibilities will be discussed in Sec.~\ref{tuning}.

\subsection{Topological phases in bands with zero Berry curvature}
We now follow a similar approach to Sec.~\ref{interacting_chern} to write down an interaction which changes the effective band structure from $K'$ to $\tilde{K}'$. We use the local-density operator $\hat{N}_\br$ as defined in Eq.~\eqref{eq:loc_dens}, but the lattice site density operator now sums over both spin species,
\be
\hat{n}_\br = \sum_{\sigma}\hat{n}_{\br\sigma}.
\ee
With this, we write
\be
U'&=&\sum_{\br\br'\sigma\sigma'}T'_{\br\br'\sigma\sigma'}  :\left[\hat{N}_{\br}c^\dagger_{\br\sigma} c^{\phantom{\dagger}}_{\br'\sigma'} +c^\dagger_{\br\sigma} c^{\phantom{\dagger}}_{\br'\sigma'}\hat{N}_{\br'}\right]:,
\ee
where any diagonal interaction-hopping terms necessary to prevent phase separation have been absorbed into $T'_{\br\br'}$. Choosing 
\be
T'_{\br\br'}&=&\tilde{t}'_{\br\br'}-t'_{\br\br'}
\ee
and following our previous reasoning, in the mean-field limit one finds
\be
K'+U'\to\tilde{K}',
\ee
which generates a spin-Chern band for each spin species. Finally, using the complete interaction $U'+\tilde{V}'$, it is possible to generate a fractional topological state from the noninteracting, zero-Berry-curvature system.

\subsection{Tuning ferromagnetism\label{tuning}}
We noted previously that Hamiltonian~\eqref{eq:HamTR} could generate either an FCI or an FTI state, depending on whether or not time reversal symmetry was spontaneously broken. For example, if the lowest two bands were filled with density $(\nu_\up,\nu_\dn)=(1/3,1/3)$, then a state with zero Chern number but nonzero spin Chern number would be produced. On the other hand, if the fillings were $(\nu_\up,\nu_\dn)=(1/3,0)$, then an FCI state with nonzero Chern number would be produced.\cite{ferromag_footnote}

These states occur at different overall fillings but also depend on whether the system is susceptible to spontaneous ferromagnetism. In order to tune this susceptibility, we introduce the spin-spin interaction
\be
V_{\mathrm{spin}}&=&\alpha\sum_\br\left(\hat{N}_{\br\up}-\hat{N}_{\br\dn}\right)^2,
\ee
where the operators
\be
 \hat N_{\br\sigma} =\frac{1}{\bar n} \sum_{\br'}  f(\br-\br' )  \hat{n}_{\br\sigma}\label{eq:loc_spin_dens}
\ee
calculate the local \emph{spin} density in a region of linear size $\xi$. For $\alpha>0$, differences in spin density will lead to an energy penalty, encouraging the fermions to be split equally between both spin species. For $\alpha<0$, spontaneous ferromagnetism will be energetically favored, and fermion spins will tend to align. By tuning $\alpha$, it is therefore possible to produce both FTI and FCI states from a topologically trivial system by adding a time-reversal-invariant interaction.

The philosophy of this section is essentially the same as the spinless fermion case: by choosing a suitable interaction, the effective band structure of the single-particle problem can be altered. Using spin-1/2 fermions allows us to do this without breaking symmetry explicitly, and we can thus generate both FCI and FTI phases. 
\section{Numerical Results\label{numerics}}
To support the arguments above, we now give some finite-size numerical results which show an FCI state that has been generated using interactions of the form described above.  Although our above arguments are rigorous in the limit of very long-range interactions, we will find that even for a short-range interaction, we can stabilize a fractional Chern insulator in a band with zero Berry curvature. 

We will consider a model of bosons hopping on a honeycomb, forming bands with zero Berry curvature (as in Sec.~\ref{interacting_chern}).    We will consider a case where the interaction strength (which determines the scale of $\tilde K$) is much greater than the hopping bandwidth $K$.  For simplicity we can set $t_{\br\br'}= K = 0$, which is appropriate in this limit.   For the kinetic energy $\tilde K$ that we wish to simulate, we consider the flattened Haldane model, which is outlined in Ref.~\onlinecite{Sheng}. The flattened Chern band hopping $\tilde{K}$ is defined by
\be
\tilde{K}_{\mathrm{HH}}&=&-\tilde{t}_1\sum_{\langle \br\br'\rangle}\left[c^\dagger_{\br'}c^{\phantom{\dagger}}_{\br}+\mathrm{H.c.}\right]-\tilde{t}_2\sum_{\langle\!\langle \br\br'\rangle\!\rangle}\left[c^\dagger_{\br'}c^{\phantom{\dagger}}_{\br}e^{i\phi_{\br\br'}}+\mathrm{H.c.}\right]\nonumber\\
&&-\tilde{t}_3\sum_{\langle\!\langle\!\langle \br\br'\rangle\!\rangle\!\rangle}\left[c^\dagger_{\br'}c^{\phantom{\dagger}}_{\br}+\mathrm{H.c.}\right] + 
M \sum_{\br} c^\dagger_{\br}c^{\phantom{\dagger}}_{\br} ,\label{eq:HH_kinetic}
\ee
where $\tilde{t}_1=1, \tilde{t}_2=0.6, \tilde{t}_3=-0.58$, $\phi=0.4\pi$, and HH stands for Haldane honeycomb. This has a flatness ratio of about 50 and exhibits a $\nu=1/2$ FCI state that remains stable even in the presence of first- and second-neighbor density-density interactions.\cite{Sheng} In order to prevent phase separation, we have also added a diagonal term with strength $M$, which in the noninteracting model appears as a chemical potential.

For simplicity, we assume we have hardcore bosons, and we take the function $f(\br)$ to be a simple top hat,
\be
f(\br'-\br)&=&\left\{\begin{array}{cc}
\frac{1}{n(l)} & \mbox{if $|\br'-\br|\leq l$}, \\
0 & \mbox{otherwise}.
\end{array}\right.
\ee
In this expression, $l$ is a length scale that covers a small number of nearest neighbors, and $n(l)$ is the number of sites that lie within the top hat, so that the function is normalized. Using Eq.~\eqref{eq:loc_dens}, we can now write the flattened Haldane model as the interaction
\be
\label{eq:UHH}
U_{\mathrm{HH}}&=&\frac{1}{2}\sum_{\br\br'}  T^{\rm HH}_{\br\br'} :\left[\hat N_\br c^\dagger_\br c^{\phantom{\dagger}}_{\br'}+c^\dagger_{\br}c^{\phantom{\dagger}}_{\br'} \hat N_{\br'}\right]:,
\ee
where $T^{\rm HH}_{\br\br'}$ contains all the hopping parameters from Eq.~\eqref{eq:HH_kinetic} (including the diagonal $M$ term). Since we are considering hardcore bosons, we do not need to add any additional interactions in order to stabilize the FCI state at half filling.

We carry out exact diagonalization of $U_{\mathrm{HH}}$ on a 24-site lattice with 6 particles and on a 32-site lattice with 8 particles. In both cases we choose the size of the top hat, defined by $l$, to extend up to third-nearest neighbors; note that this is no farther than the hopping distance. For the 24-site case we set $M=2$, and for the 32-site case we set $M=3$, chosen to give the largest many-body gap to ground-state splitting ratio. In both cases, we find strong numerical evidence to suggest that the ground state is a $\nu=1/2$ FCI. 

\begin{figure}[t]
\begin{centering}
\begin{overpic}[scale=0.35, clip=true, trim = 2 0 0 0]{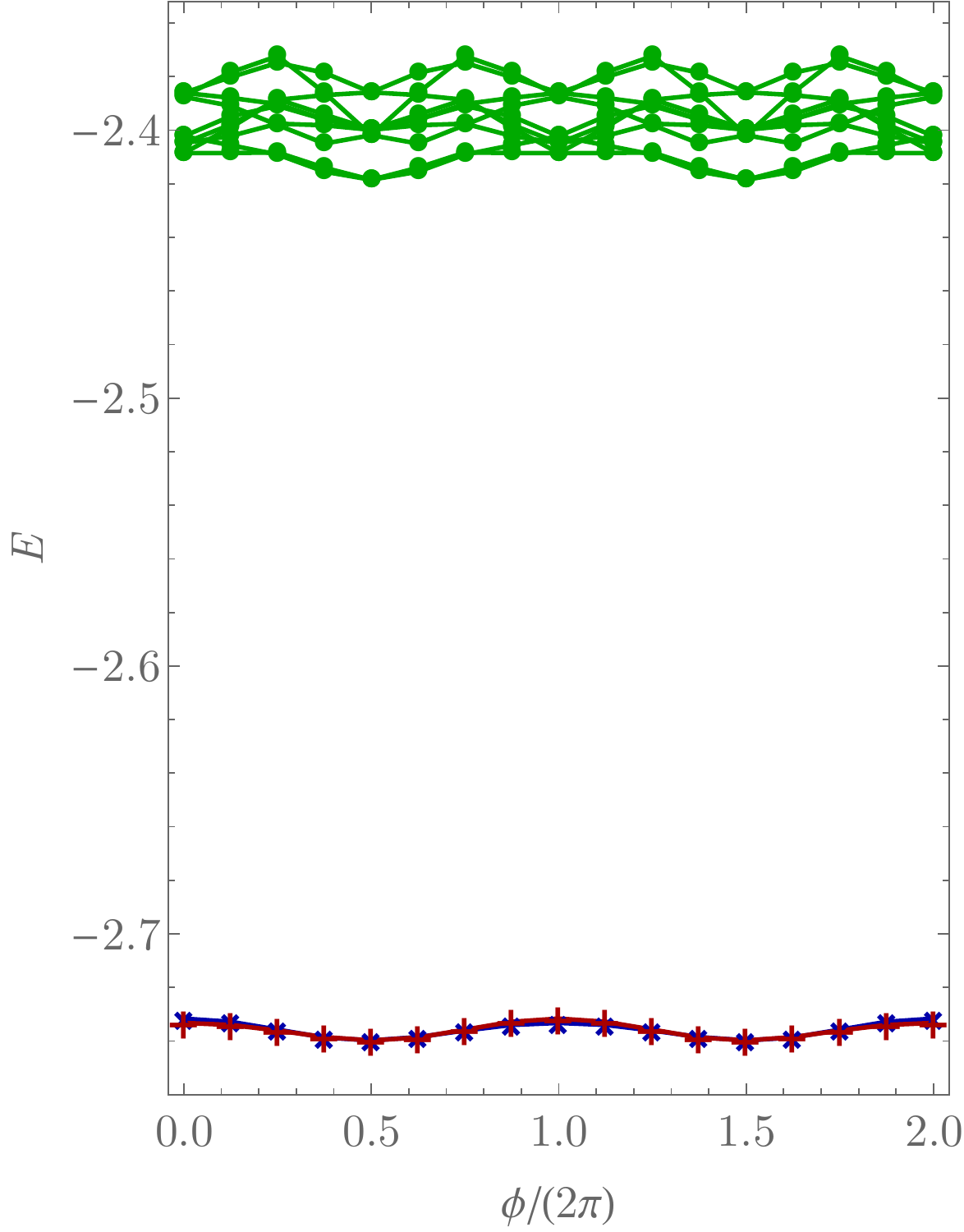} 
\put(2,51){(a)}
\put(14,34){\includegraphics[clip=true, trim = 100 200 50 150, scale=0.3]{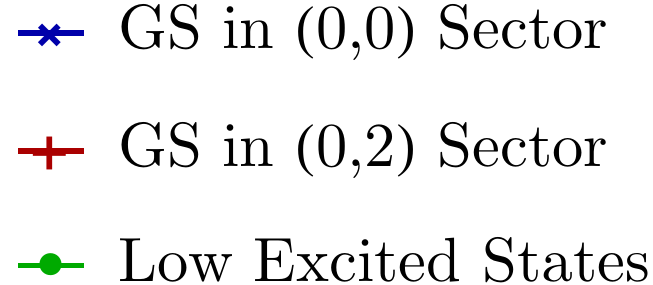}}
\put(9,11){\includegraphics[scale=0.24]{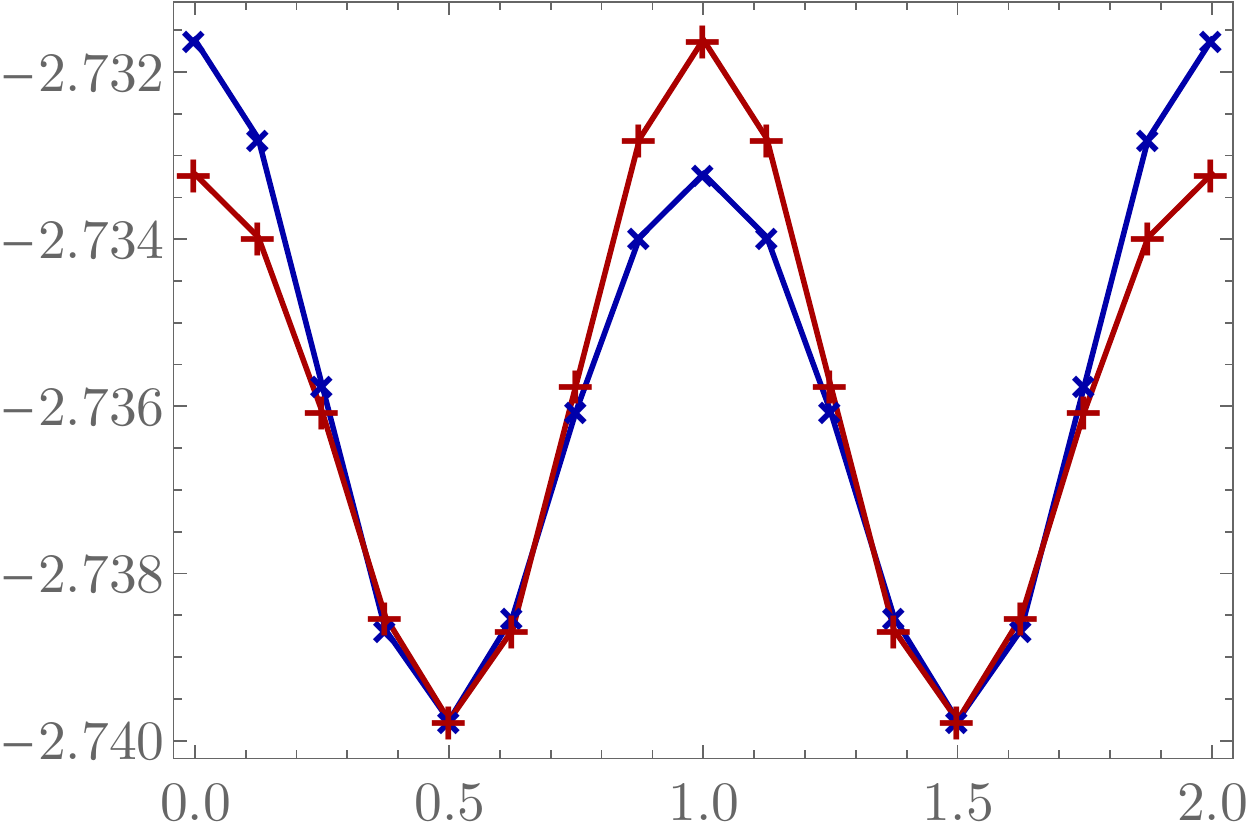} }
\end{overpic}
\begin{overpic}[scale=0.35]{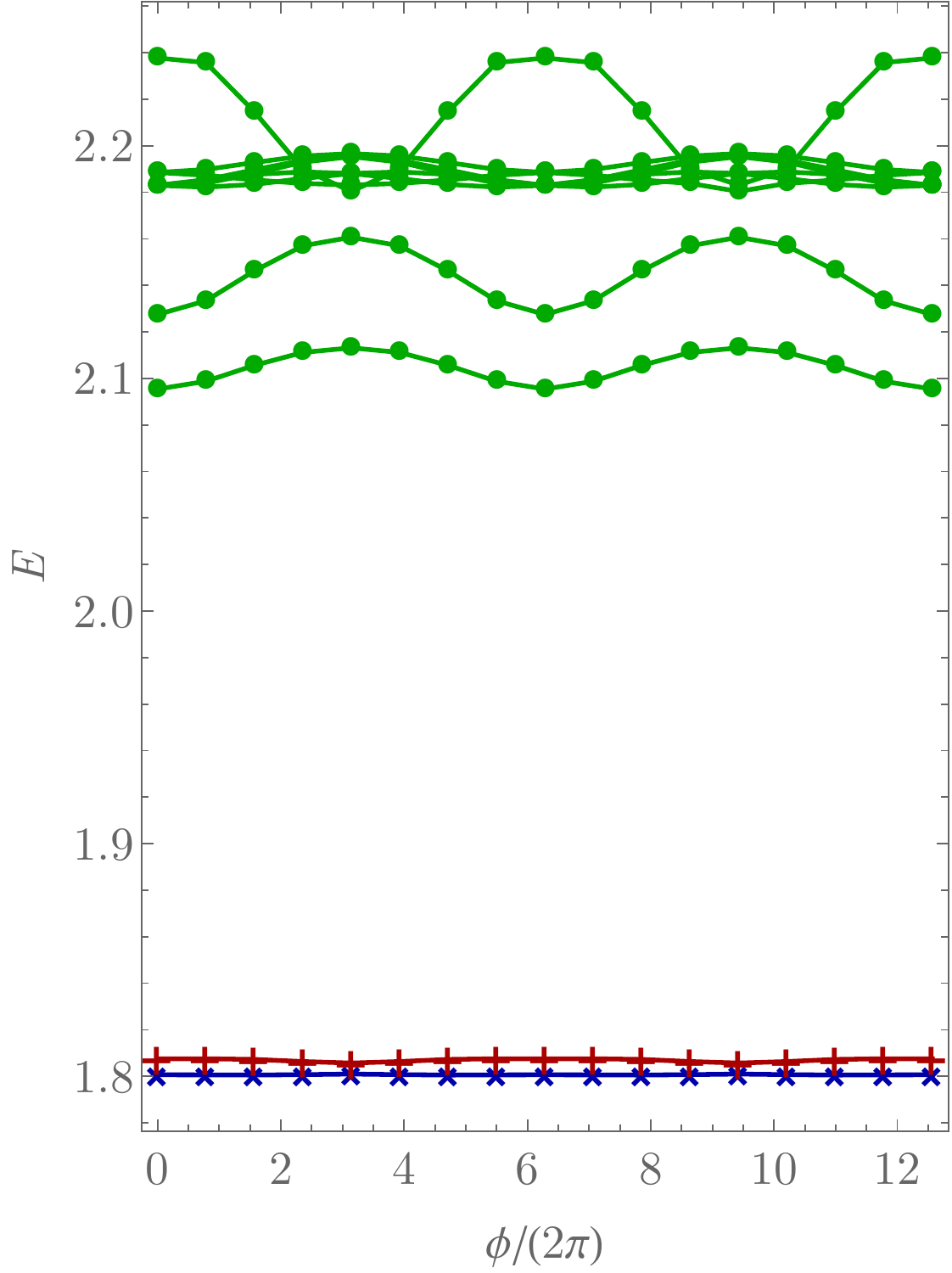} 
\put(1,51){(b)}
\put(15.5,16.5){\includegraphics[clip=true, trim = 90 200 40 150, scale=0.27]{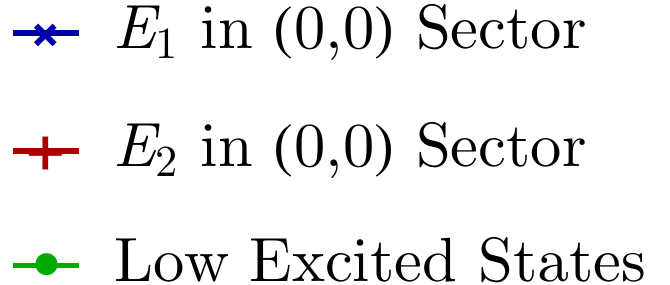}}
\put(8,11){\includegraphics[scale=0.23]{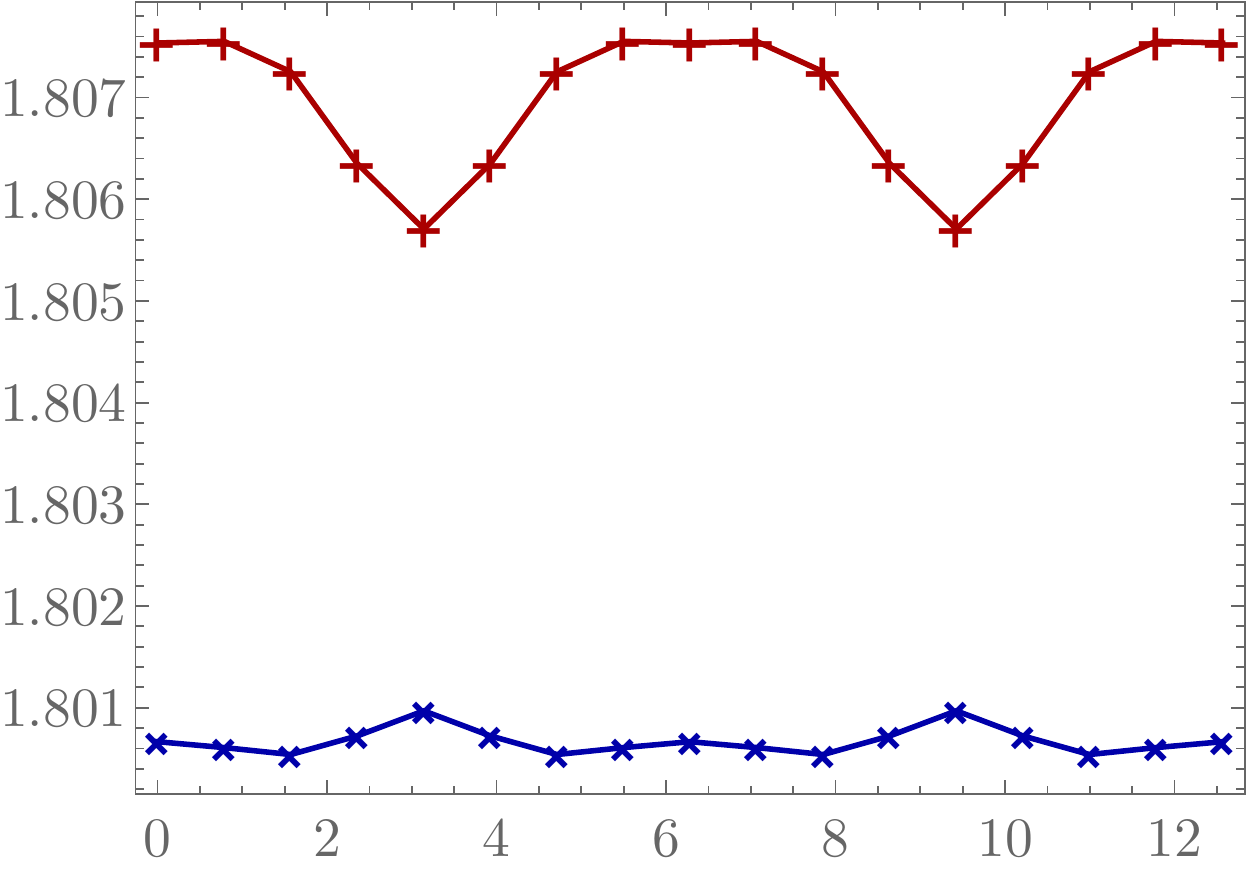} }
\end{overpic}
\caption{(Color online). Low-energy spectra as flux $\phi$ is inserted through one handle of the torus for the half-filled honeycomb lattice with (a) 24 sites and (b) 32 sites. Insets show a close-up of the ground state energy splitting. This figure gives evidence that the ground state of our model Hamiltonian [bosons on a honeycomb with zero hopping and interaction given by Eq. \ref{eq:UHH}] is a fractional Chern insulator. See main text for details of the model. \label{fluxthread}}
\end{centering}
\end{figure}

In Fig.~\ref{fluxthread} we plot the low-lying energy levels for each lattice size as magnetic flux $\phi$ is threaded through one handle of the torus. In each case we find a twofold-degenerate ground state that is separated from excited states by a large many-body gap and that occurs at the momentum values predicted for a $\nu=1/2$ FCI by the generalized Pauli principle.\cite{BernevigCounting} As flux is threaded through the handles of the torus, the two ground states for the 24-site case evolve into each other with a level crossing. For the 32-site case, the two ground states are in the same momentum sector, and display an avoided crossing under a flux insertion. 

\begin{figure}[t]
\begin{centering}
\begin{overpic}[scale=0.47, clip=true, trim = 2 0 0 0]{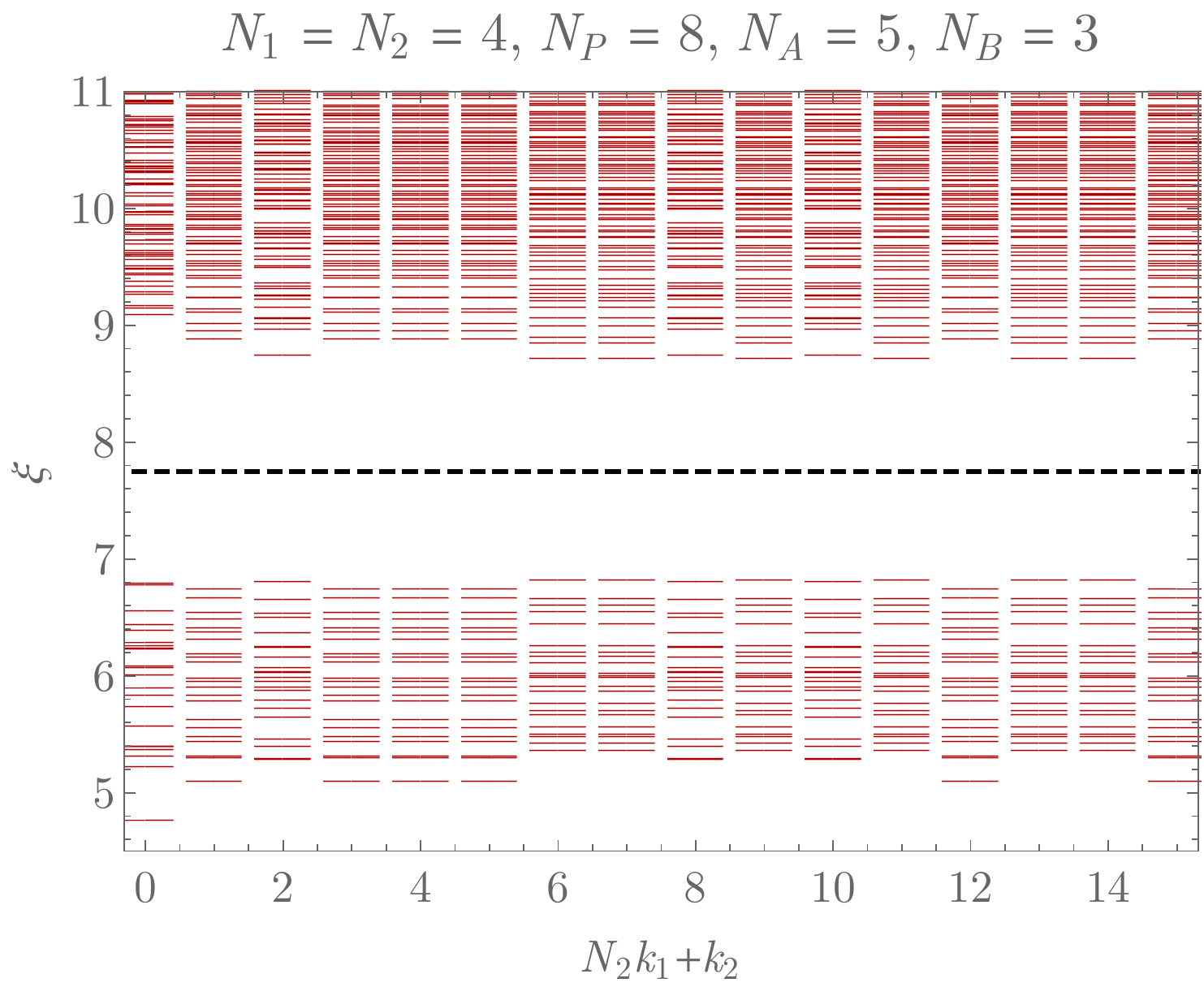} 
\end{overpic}
\caption{(Color online). Particle entanglement spectrum for $N=8$ hardcore bosons on a 32-site honeycomb, with $N_B=3$ particles traced out. $N_1$ and $N_2$ are the numbers of unit cells in each direction, with $k_1$ and $k_2$ being the corresponding momentum integers in the Brillouin zone. The number of states below the dashed line matches the expected quasihole counting rules (22 states in each sector, giving a total of 352 states.). This figure gives evidence that the ground state of our model Hamiltonian [bosons on a honeycomb with zero hopping and interaction given by Eq.~\ref{eq:UHH}] is a fractional Chern insulator.   See main text for details of the model.  \label{pes}}
\end{centering}
\end{figure}

We have also calculated the particle entanglement spectrum for various partitions for each lattice size.\cite{LiHaldane, SterdyPRL} In Fig.~\ref{pes} we show the particle entanglement spectrum for the 32-site lattice, tracing out three of the eight particles (we form the density matrix from an incoherent superposition of the two degenerate ground states, $\hat{\rho}=\frac{1}{2}\sum\ket{\psi_i}\bra{\psi_i}$, as motivated in Ref.~\onlinecite{SterdyPRL}). The entanglement spectrum shows a clear entanglement gap, and the number of states below this gap obeys the expected $(1,2)$-admissible quasihole counting rules for each momentum sector.\cite{BernevigCounting} This strongly suggests that our interaction-stabilized state is in the same universality class as other $\nu=1/2$ FQH and FCI states.

\section{Discussion}
In this paper we have proposed a type of interaction that may be used to generate fractional topological phases from a topologically trivial band structure. The underlying idea throughout is that hopping, i.e. kinetic-energy terms, are effectively a subset of the possible interaction terms.  As such, it is possible to change the single-particle spectrum using an interaction rather than a kinetic-energy term. We have outlined how this may be achieved by interacting with a local density in a region of size $\xi$: it is possible that other approaches exist which use the same philosophy.

Although we have used a formally short-range expression for $\hat{N}_\br$ in the arguments above, our approach should also apply more generally (including, for example, cases where $\hat{N}_\br$ decays algebraically). The essential requirement is that the function $f(\br'-\br)$ be normalized as in Eq.~\eqref{eq:f_normalization}, although it will also need to be suitably well behaved so that it does not favor charge-density-wave states.\cite{referee_footnote}

We note, however, that to change the Chern number of a band using an interaction, the interaction must be strong enough to mix the single-particle bands; that is, it must be on the order of the gap between the bands before the interaction is added. It is often the case that FCIs are studied for flat bands in the very weak interaction limit.  For this case our strategy will not work since a very weak interaction would not be able to change the effective Chern number of the band. However, one could also consider a more complicated strategy where the lattice translational symmetry is broken down further, enlarging the unit cell such that a band with zero Chern number is broken into two bands with opposing nonzero Chern numbers. This could be done with an arbitrarily weak perturbation.

Although the interactions we consider are not of the density-density type, density-dependent hopping interactions are predicted to arise in systems of cold atoms under certain conditions. They may arise as higher-order terms in certain tight-binding approximations\cite{Hirsch} and are also suggested to occur in effective Hamiltonians when band mixing is important.\cite{DuanEffectiveHam,KestnerEffectiveHam} More directly, density-assisted hopping terms may be induced using a periodically driven magnetic field to modulate interaction strengths,\cite{RappBosonDensity,LibertoFermionDensity} an approach which may also be used to produce density-dependent synthetic gauge fields.\cite{GreschnerGaugeDensity} In addition, \emph{vacancy}-assisted hoppings have recently been discussed in the context of kinetically constrained models, where they may also lead to topologically ordered phases.\cite{Kourtis2015} 

Nonetheless, these proposals have yet to be realized in a laboratory, and in their current formulation would not be able to realize the specific local density Hamiltonians discussed in this paper. For the time being, experimental realizations of FCI states seem most likely to be achieved through the conventional Chern band route, a direction in which substantial progress has already been made.\cite{Aidelsburger2013,Miyake2013,Jotzu2015}

Finally, it is interesting to note that our construction provides a nice counterexample to the claims of Ref.~\onlinecite{Chamon} (which have also been disproven in another manner by the current authors in Ref.~\onlinecite{Us}).  Reference~\onlinecite{Chamon} claims to calculate the Hall conductivity of an FCI as an integral of the Berry curvature of the noninteracting bands multiplied by some occupancies over the Brillouin zone.  It is clear that this cannot be correct since in our construction the Berry curvature of the noninteracting bands is strictly zero (so that Ref.~\onlinecite{Chamon} would always predict zero Hall conductivity) whereas, due to the (time-reversal-breaking) interaction, the system is an FCI with quantized, nonzero, Hall conductivity.

\vspace*{10pt}

\begin{acknowledgments}
We are grateful for useful discussions with A. Grushin and J. Motruk. S.H.S. and F.H. are supported by EPSRC Grants No. EP/I032487/1 and No. EP/I031014/1. N.R. is supported by NSF Grant No.~DMR-1408916. Statement of compliance with EPSRC policy
framework on research data: This publication reports theoretical work that does not require supporting research data.
\end{acknowledgments}

\appendix
\section{Berry Curvature and Berry Matrix\label{berry_def}}
In this appendix we will briefly outline the definitions of Berry curvature that we use in the main text. First, we write the single-particle Bloch solution for the $a$th band as
\be
\psi_{a\bk}(\br)&=&e^{i\bk\cdot\br}u_{a\bk}(\br),
\ee
where $u_{a\bk}(\br+\bR)=u_{a\bk}(\br)$ is periodic under lattice translations. For a single band, the (gauge-invariant) Berry curvature is defined through 
\be
F^a(\bk)&=&i\bigg\langle\frac{\partial u_{a\bk}}{\partial\bk}\bigg|\times\bigg|\frac{\partial u_{a\bk}}{\partial\bk}\bigg\rangle,
\ee
where, for a two-dimensional system in the $xy$ plane, we mean the $k_z$ component of the cross product. Integrating this quantity over the Brillouin zone (and dividing by $2\pi$) gives the Chern number, which is proportional to the transverse Hall conductivity of the filled band and must take an integer value.

When there are several degenerate (or nearly degenerate) bands, the generalization of the Berry curvature is the Berry matrix,
\be
F^{ab}(\bk)&=&i\bigg\langle\frac{\partial u_{a\bk}}{\partial\bk}\bigg|\times\bigg|\frac{\partial u_{b\bk}}{\partial\bk}\bigg\rangle\\
&&+i\sum_{c}\bigg\langle u_{a\bk}\bigg|\frac{\partial u_{c\bk}}{\partial\bk}\bigg\rangle\times\bigg\langle u_{c\bk}\bigg|\frac{\partial u_{b\bk}}{\partial\bk}\bigg\rangle,\nonumber
\ee
where the sum over $c$ includes all bands in the degenerate multiplet. The Berry matrix is not gauge invariant itself, but measurable quantities are contained within gauge-invariant expressions like the trace. We note that the components of this matrix vanish trivially if the multiplet contains all bands of the system.

In an FCI state, the noninteracting band picture breaks down and we can no longer calculate the Chern number from the single-particle properties. Instead, the many-body Berry curvature is calculated by threading flux through the handles of the underlying torus. If the many-body ground state is written as $\ket{\Psi}$ and we thread flux $\phi_1$ and $\phi_2$ through each handle, then the many-body Berry curvature is defined through
\be
F(\phi_1,\phi_2)&=&i\left[\bigg\langle\frac{\partial \Psi}{\partial\phi_1}\bigg|\frac{\partial \Psi}{\partial\phi_2}\bigg\rangle-\bigg\langle\frac{\partial \Psi}{\partial\phi_2}\bigg|\frac{\partial \Psi}{\partial\phi_1}\bigg\rangle\right].
\ee
The many-body Chern number can then be calculated by integrating the above quantity over $\phi_1$ and $\phi_2$.
\section{Stability Beyond Mean-Field Theory\label{pert_stab}}
In Sec.~\ref{interacting_chern} we introduced an interaction that reproduces the kinetic energy of a target Chern band model exactly in the mean-field limit. In this appendix we will consider perturbative corrections to this mean-field picture and justify the stability of the FCI ground state. 

When the correlation length $\xi$ tends to infinity (or reaches the system size in the finite-size case), the ``local'' density operator $\hat{N}_\br$ can be replaced with unity. In general, the operator $U$ can be written as $U=U_{\mathrm{MF}}+\delta U$, with
\be
U_{\mathrm{MF}}&=& \sum_{\br\br'}   T_{\br\br'} c^\dagger_{\br} c^{\phantom{\dagger}}_{\br'},\\
\delta U&=& \frac{1}{2} \sum_{\br\br'}   T_{\br\br'} :\left[\delta\hat N_\br c^\dagger_{\br} c^{\phantom{\dagger}}_{\br'}+c^\dagger_{\br} c^{\phantom{\dagger}}_{\br'}\delta\hat N_{\br'}\right]:,
\ee
and 
\be
\delta\hat{N}_\br=\hat{N}_\br-1=\frac{1}{\bar{n}}\sum_{\br'}f(\br'-\br)\left[\hat{n}_\br-\bar{n}\right].
\ee
In this way, the perturbation $\delta U$ depends on the density fluctuations in a region of linear dimension $\xi$, scaled by the average particle density $\bar{n}$.

In order to estimate the size of the corrections due to $\delta U$, we first recall that for a general, uncorrelated Poisson process, density fluctuations in an area of linear size $\xi$ scale as $\delta\hat{N}_\br\sim1/\xi$. Quantum Hall states are more correlated than this, so we expect their density to fluctuations to scale as $\delta\hat{N}_\br\sim1/\xi$ \emph{at most}. Assuming that this property transfers to systems on a lattice, we expect $\delta N_\br$ acting on an FCI ground state to have an effect that is no larger than $O(1/\xi)$. 

The action of the hopping component of $\delta U$ is bounded by the maximum bandwidth $W$, multiplied by the number of neighboring sites and orbitals involved in the sum $h$. At $p$th order in perturbation theory, the action of $\delta U$ on the FCI ground state is then bounded by
\be
\delta U^{(p)}&<&\left(A\frac{hW}{\xi}\right)^p,
\ee
with $A$ being a constant that depends on energy denominators and scale factors of order unity. By choosing $\xi$ large enough, we can ensure that perturbative corrections of this form are negligible and convergent.

However, our assumption that $\delta\hat{N}_\br\sim1/\xi$ is valid only at low orders in perturbation theory when the state being acted upon is close to a (homogeneous) FCI ground state. After applying $\delta U$ many times, perturbative hopping processes may generate virtual phase-separated states, which have $\delta\hat{N}_\br\sim1$. Assuming that the hopping described by $T_{\br\br'}$ can transfer particles by up to $L$ sites, virtual states that move particles by linear dimension $\xi$ will first start to appear at order $\xi/L$ in perturbation theory.   A finite  density of particles over a region of area $\xi^2$ may then be moved, hence significantly changing the value of $\hat N$ at order $(\xi/L)^{C \xi^2}$ in perturbation theory, where $C$ is some constant.  We can push these contributions to arbitrarily high order in perturbation theory by making $\xi$ larger. Additionally, we can suppress these perturbative corrections with a large energy denominator by choosing $M$, the diagonal interaction term in Eq.~\eqref{modified_T}, to be large and by adding additional density-density interactions between nearby neighbors.  Finally, we note that our numerical work confirms that even for small $\xi$, fractional Chern insulators may be stabilized.



\begin{thebibliography}{1}
\bibitem{ordering_footnote} The normal ordering is important if $U$ is to be a pure interaction: if they are not normal ordered, terms of the form $c^\dagger_{\br'} c^{\protect\phantom{\dagger}}_{\br}c^\dagger_{\br}c^{\protect\phantom{\dagger}}_{\br}$ will contain pieces that may be written as pure hopping terms using anticommutation relations.
\end{thebibliography}
\end{document}